\documentclass[
showpacs,preprintnumbers,
nofootinbib,
 amsmath,amssymb,
 aps,
prd,
twocolumn
]{revtex4-1}

\usepackage[utf8]{inputenc}
\usepackage{epsf}
\usepackage{graphicx} 
\usepackage{amssymb}
\usepackage{dcolumn}
\usepackage{amsmath}
\usepackage{hyperref}
\usepackage{multirow}
\usepackage{booktabs}
\usepackage{siunitx}
\newcommand{\kcmb}{\kappa_{\rm cmb}}
\newcommand{\kgal}{\kappa_{\rm gal}}
\def\thetaB{\mbox{\boldmath$\theta$}}

\def\ellB{\mbox{\boldmath$\ell$}}

\def\lsim{~\rlap{$<$}{\lower 1.0ex\hbox{$\sim$}}}
\def\gsim{~\rlap{$>$}{\lower 1.0ex\hbox{$\sim$}}}
\def\CB{\mbox{\boldmath${\rm C}$}}

\def\apj{ApJ\,}                 
\def\apjl{ApJL\,}                
\def\apjs{ApJS\,}               
\def\mnras{MNRAS\,}             
\def\aap{A\&A\,}                
\def\physrep{Phys.~Rep.\,}   

\def\jcap{JCAP\,}

\begin{document}

\title{Cross-correlation of Planck CMB Lensing\\ and CFHTLenS Galaxy Weak Lensing Maps}
\author{Jia Liu}
 \email{jia@astro.columbia.edu}
\author{J. Colin Hill}
 \email{jch@astro.columbia.edu}
\affiliation{Department of Astronomy and Astrophysics,\\ Columbia University, New York, NY 10027, USA
}

\date{\today}
\begin{abstract}

We cross-correlate cosmic microwave background (CMB) lensing 
and galaxy weak lensing maps using the Planck 2013 and 2015 data 
and the 154 deg$^2$ Canada-France-Hawaii Telescope Lensing Survey (CFHTLenS). This
measurement probes large-scale structure at intermediate redshifts $\approx 0.9$,
between the high- and low-redshift peaks of the CMB and CFHTLenS lensing kernels,
respectively. Using the noise properties of these data sets and standard Planck 2015 $\Lambda{\rm CDM}$ cosmological
parameters, we forecast a signal-to-noise ratio $\approx 4.6$ for the cross-correlation.
We find that the noise level of our actual measurement agrees well with this estimate, but the amplitude of the 
signal lies well below the theoretical prediction. The best-fit amplitudes of our measured cross-correlations are $A_{2013}=0.48
\pm0.26$ and $A_{2015}=0.44\pm0.22$ using the 2013 and 2015 Planck CMB lensing maps, respectively, where
$A=1$ corresponds to the fiducial Planck 2015 $\Lambda{\rm CDM}$ prediction.
Due to the low measured amplitude, the detection significance is moderate ($\approx 2\sigma$)
and the data are in tension with the theoretical prediction ($\approx 2$--$2.5\sigma$). The tension is
reduced somewhat when compared to predictions using WMAP9 parameters, for which we find
$A_{2013}=0.56 \pm0.30$ and $A_{2015}=0.52\pm0.26$.  We consider various systematic
effects, finding that photometric redshift uncertainties, contamination by intrinsic alignments, and effects due to the masking of galaxy
clusters in the Planck 2015 CMB lensing reconstruction are able to help resolve the tension at a significant level ($\approx 10\%$ each).
An overall multiplicative bias in the CFHTLenS shear data could also play a role, which can be
tested with existing data.  We close with forecasts for measurements of the CMB lensing -- galaxy lensing
cross-correlation using ongoing and future weak lensing surveys, which will definitively test the significance of the tension
in our results with respect to $\Lambda{\rm CDM}$.
\end{abstract}

\pacs{98.80.-k, 98.62.Sb, 98.70.Vc}

\maketitle

\section{Introduction}\label{sec:intro}

Gravitational lensing by large-scale structure is a promising cosmological probe.
During their cosmic journey toward us, photons emitted at cosmological distances
are deflected by the intervening matter. As a result, we see a distorted image
of the source light distribution.
Lensing distortions produce non-gaussianity in maps of cosmic microwave background (CMB)
temperature and polarization anisotropies. Lensed galaxies are magnified in brightness and 
weakly distorted (sheared) from their intrinsic shape. Statistical measurements of
CMB lensing~\cite{Das2011, vanEngelen2012, Planck2013XVII} and galaxy
weak lensing~\cite{Schrabback2010, Heymans2012} have been achieved
recently, and are now a useful tool for precision cosmology~\cite{Planck2015XV}.

The cross-correlation of CMB lensing maps with other tracers of large-scale structure
can provide additional cosmological and astrophysical information. For example,
cross-correlations with galaxy or quasar density maps measure the bias of the objects~(e.g.,
\cite{Sherwin2012,Bleem2012,Geach2013,Allison2015,Omori2015}), while cross-correlations with
cosmic infrared background (CIB) or thermal Sunyaev-Zel'dovich (tSZ) effect maps provide information
on the complex relationship between the dark matter and the baryons in different forms
over cosmic time (e.g., dusty star-forming galaxies or hot, ionized gas)~\cite{Planck2013XVIII,Holder2013,vanEngelen2014,Hill2014}.

Similarly, cross-correlating CMB lensing and galaxy weak lensing maps can provide useful
cosmological information. While CMB lensing and current galaxy lensing surveys are most 
sensitive to matter fluctuations at different redshifts~($z\approx1$--$2$ and $z \lesssim 0.5$, 
respectively), their cross-power spectrum is sensitive to large-scale structure at intermediate
redshifts $z\approx0.9$.
Combining the auto- and cross-power spectra can thus
provide tomographic information on the growth of structure.
Furthermore, the cross-power spectrum is immune to nearly all systematic effects that can
plague measurements of the lensing convergence auto-power spectrum (e.g., the point spread
function (PSF) correction, for galaxy shapes), since the CMB and galaxy lensing surveys are
completely independent measurements. In fact, the CMB lensing -- galaxy lensing cross-correlation
can be used to measure the multiplicative bias in galaxy lensing shear maps, thus overcoming
an important systematic in cosmic shear analyses~\cite{Vallinotto2012,Das2013}.

Ref.~\cite{Hand2013} (H15) reported the first detection of the cross-correlation of CMB 
lensing and galaxy lensing with a significance of 4.2$\sigma$, using CMB lensing maps
from Atacama Cosmology Telescope~(ACT) data and galaxy lensing maps from the
Canada-France-Hawaii Telescope Stripe 82 Survey~(CS82). They found best-fit 
amplitudes $A=0.78\pm0.18$ with respect to a fiducial model based on Planck 2013 cosmological parameters, and 
$A=0.92\pm0.22$ for a model based on
Wilkinson Microwave Anisotropy Probe~(WMAP) parameters.
They also noted that uncertainty in the redshift distribution of their source galaxies, determined from cross-matched COSMOS redshifts for a small subset of the data,
could cause $10-20\%$ changes in the theoretical prediction.

In this work, we perform a similar analysis using CMB lensing maps from the Planck 
satellite (2013 and 2015 data releases)\footnote{\url{http://www.cosmos.esa.int/web/planck}} and galaxy weak lensing data from the 
Canada-France-Hawaii Telescope Lensing Survey~(CFHTLenS)\footnote{\url{http://www.cfhtlens.org/}}. CFHTLenS
has a similar survey size and depth as CS82, and the Planck 2015 CMB lensing reconstruction
noise is comparable to that in the ACT lensing reconstruction used in H15
(but with somewhat different $\ell$-dependence due to the different resolutions
of the two experiments).  Therefore, we expect our detection to be of comparable significance to that found
in H15.  Moreover, since the Planck CMB lensing map covers nearly the full sky, the outlook for cross-correlations of these data
with ongoing wide-field galaxy lensing surveys (e.g., the Dark Energy Survey\footnote{\url{http://www.darkenergysurvey.org/}} and Hyper Suprime-Cam Survey~\footnote{\url{http://www.naoj.org/Projects/HSC/}}) is promising.
We make predictions for these surveys, and also compare our cross-correlation results between the two Planck data releases.

This paper is structured as follows. We first
introduce the lensing formalism in Sec.~\ref{sec: formalism}, and describe our data
analysis in Sec.~\ref{sec: data}. We then present our results in 
Sec.~\ref{sec: results} and summarize in Sec.~\ref{sec: summary}.

\section{Formalism}\label{sec: formalism}

The lensing convergence is a weighted projection of the three-dimensional matter overdensity $\delta=\delta\rho/\rho$ along the line of sight,

\begin{equation}
\kappa(\thetaB) = \int_0^{\infty} dz W(z) \delta(\chi(z)\thetaB, z),
\end{equation}
where $\chi(z)$ is the comoving distance to redshift $z$ and the kernel $W(z)$ indicates the lensing strength  at redshift $z$ for sources with a redshift distribution $dn(z)/dz$. For a flat universe,
\begin{eqnarray}
W(z) &=& \frac{3}{2}\Omega_{m} H_0^2 \frac{(1+z)}{H(z)} \frac{\chi(z)}{c} \nonumber\\
&\times&\int_z^{\infty} dz_s \frac{dn(z_s)}{dz_s} \frac{\chi(z_s) - \chi(z)}{\chi(z_s)},
\end{eqnarray}
where $\Omega_{m}$ is the matter density as a fraction of the critical density at $z=0$, $H(z)$ is the Hubble constant at redshift $z$, with a present-day value $H_0$, $c$ is the speed of light, and $z_s$ is the source redshift. Note that $\int_0^{\infty} dz_s dn(z_s)/dz_s = 1$. We hereafter denote the galaxy lensing kernel computed with the CFHTLenS source redshift distribution as $W^{\kgal}(z)$. For CMB lensing, there is only one source plane at the last scattering surface $z_\star=1100$. Using $dn(z_s)/dz_s=\delta_D(z_s-z_\star)$, where $\delta_D$ is the Dirac delta function, the CMB lensing kernel can be simplified to
\begin{eqnarray}
W^{\kcmb}(z) &=&  \frac{3}{2}\Omega_{m}H_0^2  \frac{(1+z)}{H(z)} \frac{\chi(z)}{c} \nonumber\\ 
&\times&  \frac{\chi(z_\star)-\chi(z)}{\chi(z_\star)}.
\end{eqnarray}

\begin{figure}
\begin{center}
\includegraphics[scale=0.45]{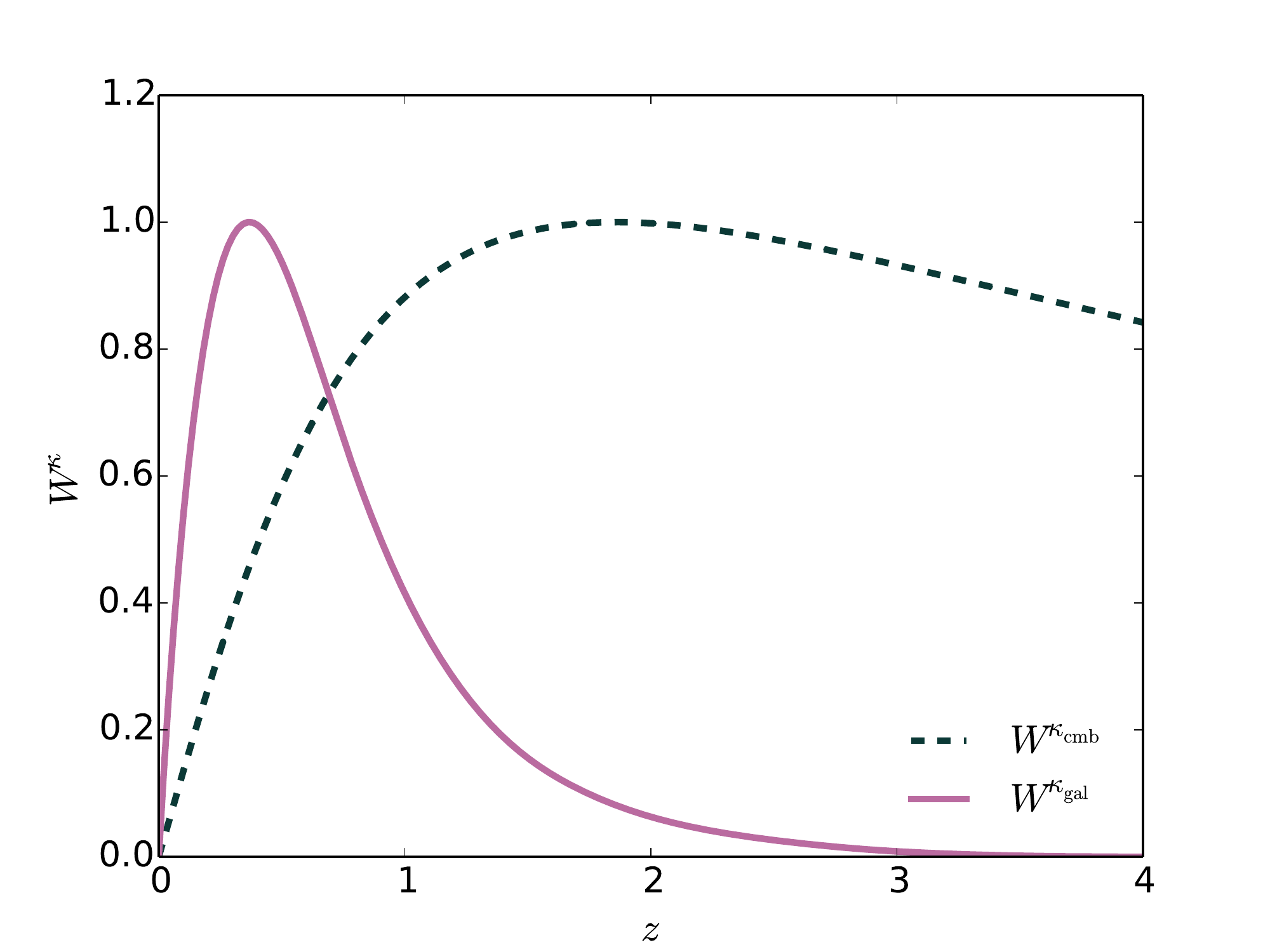}
\caption{\label{fig: lensing_kernels} The lensing kernels for the CMB (dashed curve) and CFHTLenS galaxies (solid curve), normalized to a maximum value of unity.}
\end{center}
\end{figure}

The lensing kernels for the CMB and CFHTLenS galaxies are shown in Fig.~\ref{fig: lensing_kernels}. We discuss the CFHTLenS source distribution in detail in the next section.
The mean redshift weighted by the two lensing kernels is $z_{\rm mean}=\int_0^{\infty} W^{\kcmb}W^{\kgal}z dz/\int_0^{\infty} W^{\kcmb}W^{\kgal} dz \approx0.9$.

In the Limber approximation \cite{Limber1954}, the CMB lensing-galaxy lensing cross-correlation is

\begin{equation}\label{eqn:theory_eqn}
C_\ell^{{\kcmb}{\kgal}} = \int_0^\infty \frac{dz}{c} \frac{H(z)}{\chi(z)^2} W^{\kcmb}(z)W^{\kgal}(z) \\
P\left(k, z \right),
\end{equation}
where $P\left(k, z \right)$ is the matter power spectrum evaluated at wavenumber $k = \ell/\chi(z)$ at redshift $z$.
For our fiducial theoretical calculations, we compute Eq.~\ref{eqn:theory_eqn} with $P\left(k, z \right)$ from the code {\texttt nicaea}\footnote{\url{http://www.cosmostat.org/software/nicaea/}}, using the nonlinear power spectrum from HALOFIT~\cite{Smith2003, Takahashi2012}. For a comparison, we also compute
theoretical predictions using the halo model (e.g.,~\cite{Seljak2000,Cooray2002}) following the methodology described in~\cite{Hill2014,Battaglia2014} (we simply replace the tSZ signal in their approach with the CFHTLenS galaxy lensing signal). Since the halo model is only expected to be accurate to $5$-$10$\% precision in this context, we use the {\texttt nicaea}+HALOFIT approach when comparing our measurements to theory. However, the halo model calculation provides intuition about the influence of nonlinear power, as it explicitly separates the one-halo and two-halo contributions to the cross-power spectrum. Finally, we also compute Eq.~\ref{eqn:theory_eqn} with the linear matter power spectrum from {\texttt camb}\footnote{\url{http://camb.info}} for an additional comparison.

\begin{figure}
\begin{center}
\includegraphics[scale=0.45]{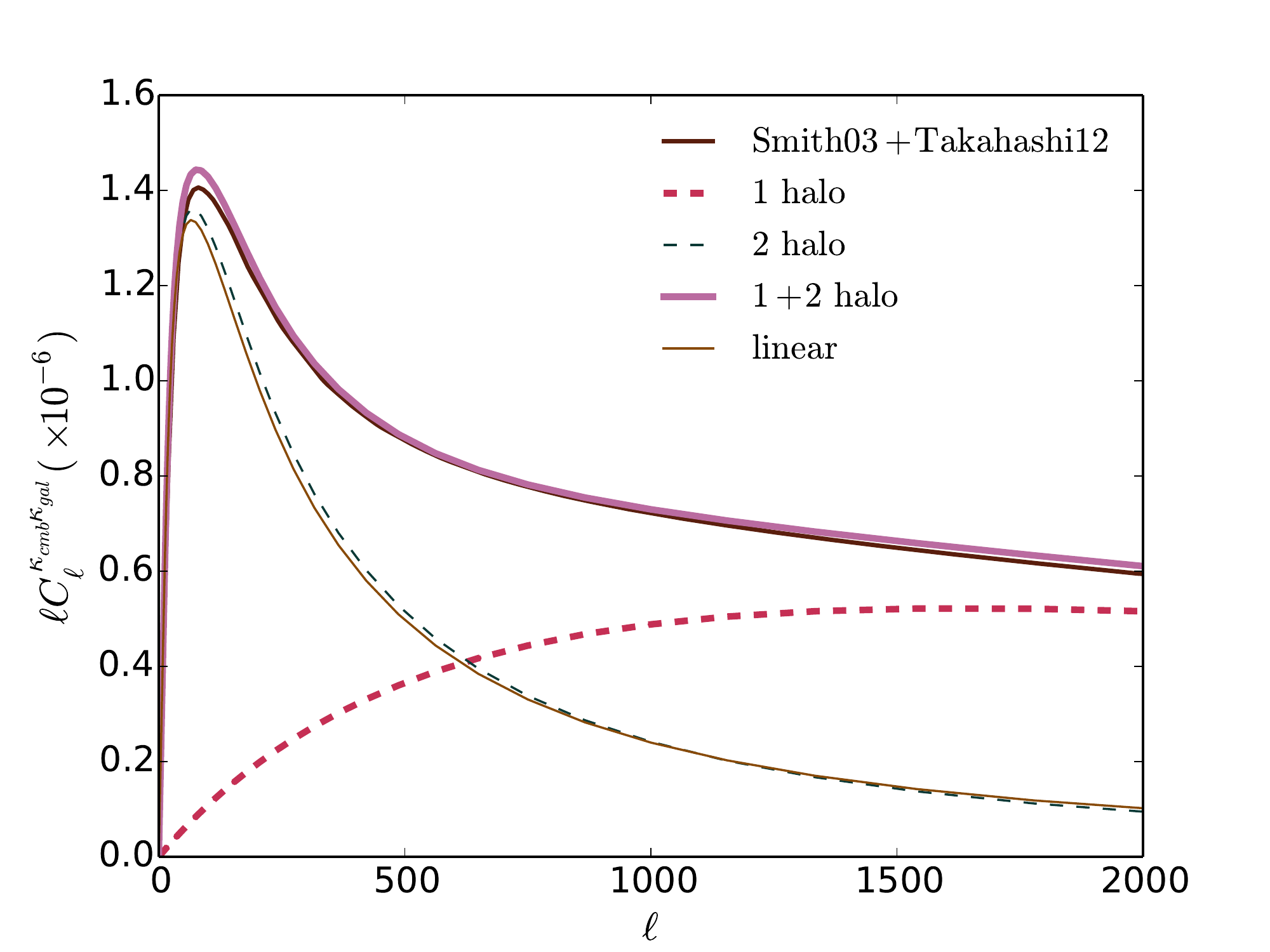}
\caption{\label{fig: model_haloterms} Theoretical predictions using Planck 2015 cosmological parameters. The thin brown solid curve labeled ``Smith03+Takahashi12" shows our fiducial theoretical calculation using {\texttt nicaea}+HALOFIT. The other curves show predictions using the halo model and the linear matter power spectrum, as labeled in the figure.}
\end{center}
\end{figure}

The predicted cross-power spectrum is shown in Fig.~\ref{fig: model_haloterms}, using Planck 2015 cosmological parameters (column 4 of Table 3 in Ref.~\cite{Planck2015XIII}). 
In particular, $\Omega_m = 0.3156$ and $\sigma_8 = 0.831$, where $\sigma_8$ is the rms amplitude of linear matter density fluctuations at $z=0$ on a scale of $8 \, h^{-1}$~Mpc. Fig.~\ref{fig: model_haloterms} shows that nonlinear contributions are non-negligible for $\ell \gtrsim 100$ and are dominant for $\ell \gtrsim 500$. Similar results are seen in the halo model comparison, where the one-halo term takes over at $\ell \approx 600$. Note that the total power predicted by the halo model is in good agreement with the more accurate HALOFIT calculation.

\begin{figure}
\begin{center}
\includegraphics[scale=0.45]{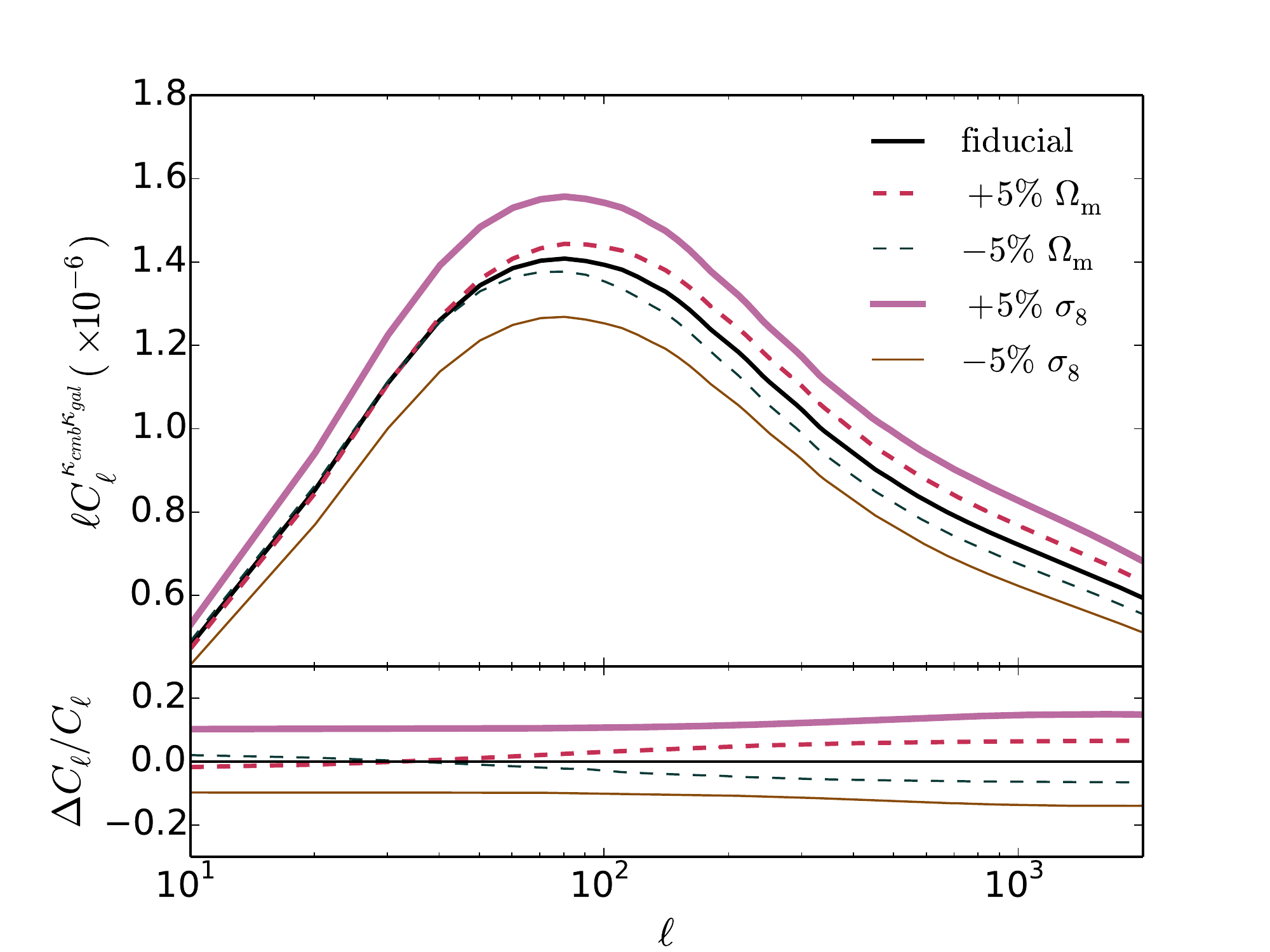}
\caption{\label{fig: model_varyingparams} Cosmological sensitivity of the cross-correlation. We fix the fiducial
cosmology at Planck 2015 cosmological parameters, and vary $\Omega_m$ and $\sigma_8$ by $\pm5\%$.}
\end{center}
\end{figure}

To demonstrate the cosmological sensitivity of the cross-power spectrum, we vary $\Omega_m$ and $\sigma_8$ by $\pm5\%$, and show the results in
Fig.~\ref{fig: model_varyingparams}. On most angular scales, the cross-power spectrum shows degeneracy
between the two parameters, where a larger (smaller) $\Omega_m$ or $\sigma_8$  
simply increases (decreases) the overall amplitude. 
However, on very large angular scales ($\ell \lesssim 30$), increasing (decreasing) $\Omega_m$ decreases (increases) the power. Thus,
in principle wide-field galaxy lensing surveys covering large sky fractions can break the degeneracy between the parameters. Over the
range of angular scales considered in this paper, the parameters are completely degenerate.

Later in the paper, we will also compare our measurements to theoretical calculations using the maximum-likelihood WMAP9+eCMB+BAO+$H_0$ 
parameters~\cite{Hinshaw2013} (see their Table 2). In this case, $\Omega_m = 0.282$ and $\sigma_8 = 0.817$.

In order to motivate our data analysis below, we consider a simple forecast for
the expected signal-to-noise ratio (SNR) of the Planck 2015 CMB lensing --
CFHTLenS galaxy lensing cross-correlation.  We use the fiducial calculation described above
with Planck 2015 cosmological parameters to compute the signal.
We compute the error bars using the analytic approximation based on the auto-power spectra
of the CFHTLenS and Planck lensing maps (e.g., Eq. 30 of Ref.~\cite{Hill2014}).
For Planck, we use the sum of the CMB lensing signal and noise power spectra provided in the 2015 data release,
while for CFHTLenS we use the measured auto-power spectrum of the
convergence maps (thus including both signal and noise as well).  The maps are described in
full detail in the next section.  Adopting the same sky fraction ($140 \, {\rm deg}^2$) and
multipole range ($40 \leq \ell \leq 2000$) as in our analysis below, we obtain a
predicted SNR $\approx 4.6$.  Since this estimate is based on the sky-averaged Planck
CMB lensing noise power spectrum rather than the actual power spectrum in the specific CFHTLenS
sky patches, the actual SNR is expected to differ slightly.  However, as
noted in Sec.~\ref{sec:intro}, this forecast is comparable to the H15 SNR $\approx 4.2$ obtained
using ACT CMB lensing maps and CS82 galaxy lensing maps.

\section{Data Analysis}\label{sec: data}

\begin{figure*}
\begin{center}
\includegraphics[scale=0.47]{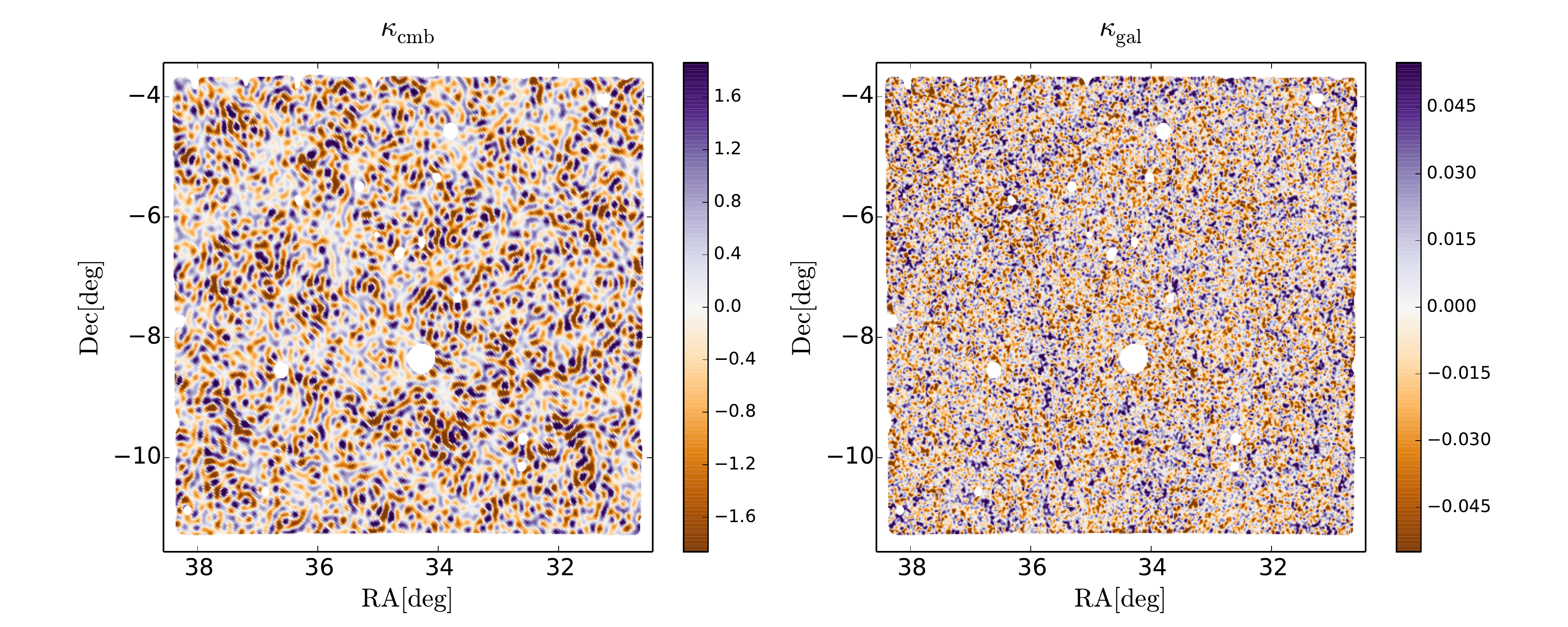}
\caption{\label{fig: sample_maps} The CMB (left) and galaxy (right) lensing maps
for the CFHTLenS W1 field. The galaxy lensing map is smoothed with a $\sigma_G = 1$ arcmin Gaussian kernel.
No filter has been applied to the CMB lensing map.
Data in the white regions are masked out due to bright point sources, such as stars in the CHFTLenS map
or radio point sources in the Planck CMB temperature maps.}
\end{center}
\end{figure*}

In this work, we use CMB lensing maps from the Planck satellite 
data releases in 2013 and 2015, and galaxy lensing maps from the CFHTLenS survey.
While Planck is a full-sky survey, CFHTLenS covers only 154~deg$^2$.
Thus, we cut out regions in the CMB lensing maps that match
the CFHTLenS fields, and construct both CMB and galaxy lensing maps in 
real space with the same resolution of 0.16 arcmin$^2$ per pixel. Fig.~\ref{fig: sample_maps}
shows examples of the CMB and galaxy lensing maps for the CFHTLenS W1 field.
We then combine the masks from both surveys and apply them to all data sets.
Finally, we analyze the cross-power spectrum between the two surveys, and 
present our results and null tests in Sec.~\ref{sec: results}.

\subsection{Planck data}\label{sec: planck}

We consider the CMB lensing maps produced by the Planck collaboration for both the 2015 and 2013 data
releases~\cite{Planck2015XV,Planck2013XVII}. Both maps are based on lensing reconstructions using
quadratic estimators~\cite{Okamoto2003}. The 2015 map~\cite{Planck2015XV} is provided as an
estimate of the CMB lensing convergence field, reconstructed using the minimum-variance combination
of all temperature and polarization estimators applied to CMB component-separated maps from
the SMICA code~\cite{Planck2015IX}. The publicly released map is band-limited to the multipole range $8 \leq \ell \leq 2048$.
We also use the associated mask, which removes regions contaminated by emission from the Galaxy and
point sources, leaving $67.3$\% of the sky. Note that the mean-field bias has already been subtracted from the publicly released map, and we thus
perform no additional such subtraction in our analysis.

For a comparison, we also consider the CMB lensing map provided in the 2013 data release~\cite{Planck2013XVII}.
In this case, the map is provided as an estimate of the CMB lensing potential $\bar{\phi}$, reconstructed using only the
temperature-based quadratic estimator applied to the 143 and 217 GHz Planck 2013 temperature maps. The
reconstruction noise levels are roughly twice as large in this map as in the 2015 map~\cite{Planck2013XVII}. 
The map is band-limited to the multipole range 
$10 \leq \ell \leq 2048$. We convert
the $\bar{\phi}$ map into a convergence map in harmonic space:
\begin{equation}\label{eqn:Planckkappaconv}
\kappa_{\rm cmb}(\ellB) = \frac{\ell(\ell+1)}{2} \frac{1}{\mathcal{R}_{\ell}^{\phi}} \bar{\phi}(\ellB),
\end{equation}
where $\mathcal{R}_{\ell}^{\phi}$ is the lensing response function provided in the 2013 data release. We then 
transform the resulting convergence map to real space in order to extract the data in the CFHTLenS regions.

We combine the associated 2013 lensing mask with the 2015 mask, although it appears
that the 2015 mask is stricter and covers essentially all of the 2013 mask, plus additional sky regions. In particular,
we note that the 2015 mask covers tSZ clusters, whereas the 2013 mask does not. The 2013 reconstruction
masks tSZ clusters in the 143 GHz channel, but not in the 217 GHz channel (where the tSZ signal is null), and thus
in the publicly released map constructed from a combination of the two channels, lensing signal is included at the location of
tSZ clusters. Since the 2015 reconstruction is based on the SMICA map, which combines all Planck
channels, tSZ clusters are masked prior to the reconstruction in order to avoid biases. We test for effects resulting from the cluster masking
in Sec.~\ref{sec: results}. We also note that biases in the Planck CMB lensing reconstruction due to tSZ or CIB leakage
should be small due to Planck's resolution and noise levels~\cite{vanEngelen2014b,Osborne2014}, even with no masking, with the
possible exception of small scales ($\ell \gtrsim 1000$) in the lensing map (however, the reconstruction noise is large on these scales).

In order to cross-correlate the Planck CMB lensing maps with the CFHTLenS convergence maps, we project the relevant regions of the CMB lensing maps (and the associated masks) onto flat-sky grids in (RA, Dec). This procedure uses a cylindrical equal-area projection implemented in the {\tt flipper} software\footnote{\url{http://www.hep.anl.gov/sdas/flipperDocumentation/}}, which was developed by members of the ACT collaboration.  The projection is performed at high resolution (HEALPix $N_{\rm side} = 8192$) in order to minimize any resulting artifacts.  We verify the accuracy of this procedure by calculating the power spectra of simulated maps before and after the projection (i.e., in the patch on the sphere and in the flat-sky projection), finding no measurable differences over the range of angular scales considered in this paper.

\subsection{CFHTLenS data}\label{sec: cfht}

The CFHTLenS survey is one of the first large galaxy lensing datasets that are publicly
available~(see also COSMOS~\cite{Scoville2007}). 
It consists of four sky patches located far from the Galactic plane, W1, W2, W3, and W4,
with a total area of 154 deg$^2$ and a limiting magnitude $i_{\rm AB}\lesssim24.5$.
The CFHTLenS data analysis pipeline consists of: (1)
creation of the galaxy catalogue using SExtractor~\cite{Erben2013};
(2) photometric redshift estimation with a Bayesian photometric
redshift code~\cite{Hildebrandt2012}; and (3) galaxy shape measurements
with {\it lens}fit~\cite{Heymans2012, Miller2013}. A summary of the
data analysis process is given in Appendix C of
Ref.~\cite{Erben2013}. We refer the reader to the CFHTLenS official papers
mentioned above for more technical details.

The procedure of our galaxy lensing map construction can be found in
Ref.~\cite{Liu2014}.  In brief, we apply a cut of \verb star_flag $\;= 0$ (requiring the object to be a galaxy), 
weight {\texttt w} $> 0$ (with larger {\texttt w} indicating smaller shear measurement uncertainty), 
and {\texttt mask} $\leq 1$ (see Table B2 in Ref.~\cite{Erben2013} for the meaning of mask values).
These cuts leave 5.3 million galaxies,
140 deg$^2$ of sky, and an 
effective number density of $n_{\rm gal} = 12.5$~arcmin$^{-2}$, with
\begin{eqnarray}
\label{eq: n_gal}
n_{\rm gal} = \frac{1}{\Omega}\frac{\left(\sum_i w_i\right)^2}{\sum_i w_i^2},
\end{eqnarray}
where $\Omega$ is the survey sky area excluding the masked regions, and $i$ denotes individual galaxies.

We then reconstruct the convergence map from shear 
measurements using \cite{Kaiser1993},
\begin{eqnarray}
\label{eq: KSI}
\hat{\kappa}_{\rm gal}(\ellB) = \left(\frac{\ell_1^2-\ell_2^2}{\ell_1^2+\ell_2^2}\right)\hat{\gamma}_1(\ellB)
+ 2\left(\frac{\ell_1\ell_2}{\ell_1^2+\ell_2^2}\right)\hat{\gamma}_2(\ellB),
\end{eqnarray}
where $\hat{\kappa}_{\rm gal}, \hat{\gamma_1}$, and $\hat{\gamma_2}$ are the 
convergence and shears in Fourier space, and $\ellB$ is the wavevector 
with components $(\ell_1, \ell_2)$. Note that in this reconstruction we correct for multiplicative
and additive biases on the shear, as given in Eqs.~(4) and~(6) in Ref.~\cite{Liu2014}. Finally, the maps are inverse Fourier-transformed into real space, and
smoothed with a $\sigma_G = 1$ arcmin Gaussian window.
Ref.~\cite{Heymans2012} identified 25\% of the 172 CFHTLenS pointings, each 
$\approx1$ deg$^2$ in size,  to have PSF residuals. Including these fields
can bias the auto-correlation function. However, we include these regions in this work,
as there is no correlation between the PSF residuals and the signal or noise in the Planck CMB lensing maps,
and the additional sky area is useful for our analysis.  Moreover, no significant change was seen in the
CFHTLenS convergence power spectrum (for $\ell < 7000$) or peak counts when including these regions in Ref.~\cite{Liu2014}.

\begin{figure}
\begin{center}
\includegraphics[scale=0.45]{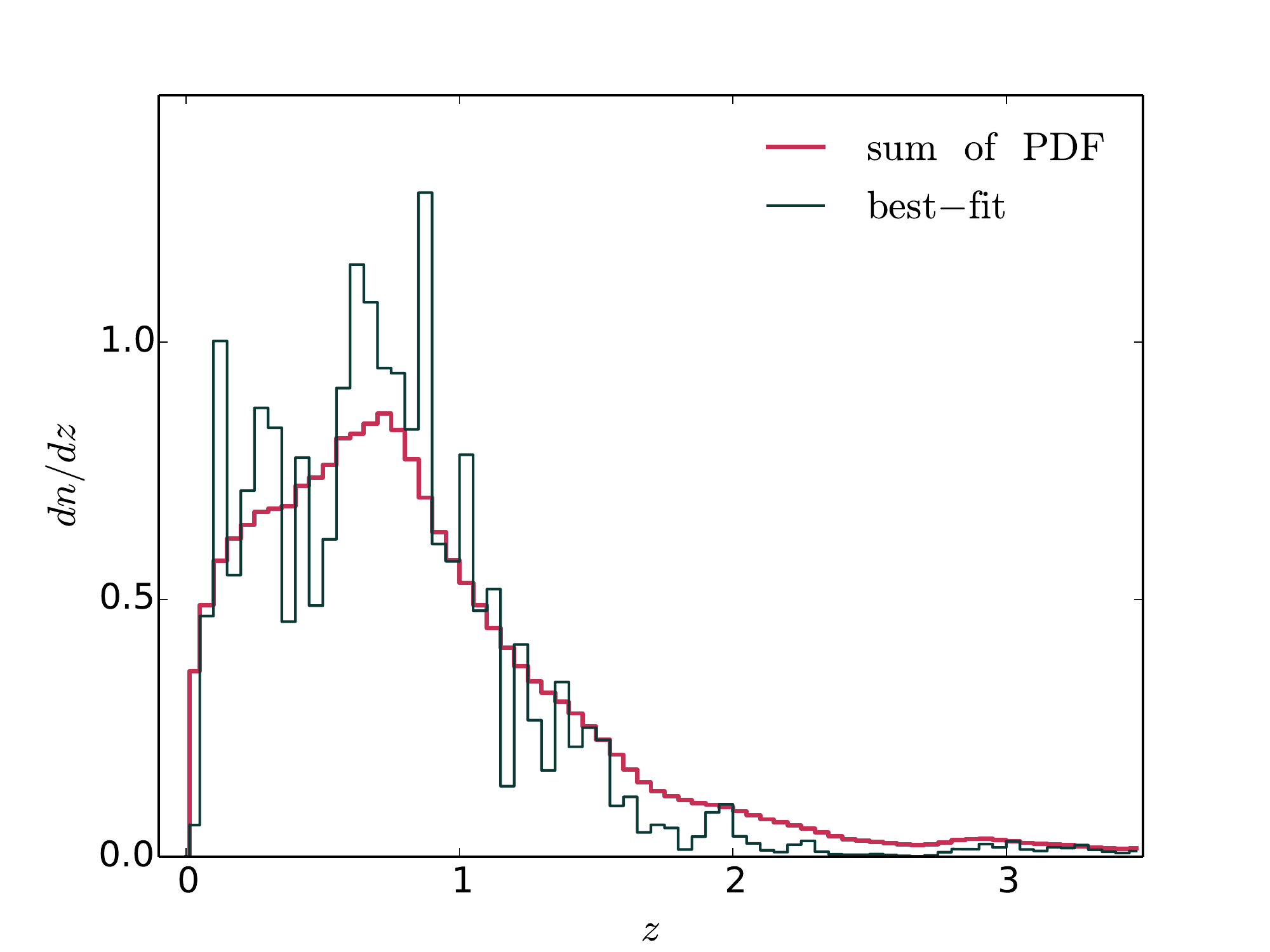}
\caption{\label{fig: dndz} CFHTLenS galaxy redshift distributions for 
the sum of PDFs of individual galaxies (red thick line) and for the best-fit 
redshift (black thin line).  We use the sum of the PDFs to model the
CFHTLenS $dn/dz$ in our analysis.}
\end{center}
\end{figure}

The redshift distributions for the source galaxies are shown in Fig.~\ref{fig: dndz},
for both the cumulative sum of the redshift probability distribution functions~(PDF) 
of individual galaxies, and the histogram of
the best-fit redshifts. We adopt the former for our analysis.
We note that we do not apply a redshift cut to the galaxy sample. 
Normally, a redshift cut of $z<1.3$ is suggested for CFHTLenS galaxies,
due to the limited number of spectroscopic redshift measurements at high-$z$ and the lack of 
a near-infrared band. At $z\approx1.3$, the $4000 \AA$ Balmer break leaves
the reddest band ($z$ band), resulting in a larger photo-$z$ uncertainty~\cite{Hildebrandt2012, Heymans2012}. 
Practically, galaxies at $z>1.3$ can 
still have high-quality shape measurements, and their lensing kernel overlaps 
more with the CMB lensing kernel. Including these galaxies thus enhances the expected
SNR of the cross-correlation signal.

To estimate the level of uncertainty due to the photometric redshifts, we first compare theoretical
models calculated using two different $dn/dz$, each computed from redshifts randomly drawn from the PDF of individual galaxies.
The resulting theoretical curves are almost identical ($<1\%$ difference). We further
investigate the potential impact from the inclusion of $z>1.3$ galaxies, which account for
$\approx 15\%$ of our total sample. As pointed out in H15, due to the strong overlap of such high-$z$ galaxies
with the CMB lensing kernel, uncertainties in their photometric redshifts can lead to non-negligible
uncertainty in the cross-correlation amplitude. Ref.~\cite{Benjamin2013} compared the summed
PDF of CFHTLenS galaxies to a matched COSMOS sample, which is measured with 30
bands and hence can be considered the ``true'' PDF, and found some discrepancies for
galaxies with $z>1.3$. 
We use the COSMOS data points from Fig.~2 of~\cite{Benjamin2013} for the highest redshift bin $z=(1.30, 7.00]$,
and replace the PDFs of our $z>1.3$ galaxies with the resulting COSMOS PDF. The theoretical model
computed using the COSMOS-corrected PDF is nearly identical to that computed using the full CFHTLenS
PDF, with only a slight decrease in the overall amplitude ($2$\%). This change is highly subdominant to
the statistical error in our measurement.

However, because COSMOS data can also suffer from systematic uncertainty in the high redshift tail (see Figs.~8 and~9 of~\cite{Ilbert2009}), 
we consider a final, crude test in which all $z>1.3$ galaxies are manually moved down to $z=1.3$. 
Under this extreme scenario, the amplitude of the theoretical model decreases by $20$\%.
However, Ref.~\cite{Ilbert2009} shows that the errors on the high-redshift photo-$z$ are approximately
symmetric, and thus it is unrealistic to expect that all such galaxies should be moved to lower redshifts.
If some were moved to higher redshifts, the theoretical amplitude would increase.
In the absence of a more precise quantifier of these uncertainties, we conclude that systematic uncertainties 
in our cross-correlation results due to photo-$z$ are on the order of $\approx 10$\%, and at most $20$\%.

\subsection{Power spectrum and covariance estimation}

We calculate the cross-power spectrum of the CMB lensing and galaxy lensing
convergence maps in the flat-sky approximation using the pipeline developed
for Ref.~\cite{Liu2014}.  First, to reduce edge effects, we mask the ten pixels nearest the edge of each map. 
We combine this mask with the Planck and CFHTLenS masks described above,
and then smooth the final mask with a $\sigma_G = 4$ arcmin Gaussian window.
We apply the apodized mask to the CMB lensing and galaxy lensing maps, and estimate the 2D power spectrum as
\begin{eqnarray}
\label{eq: ps2d}
C^{\kappa_{\rm cmb}\kappa_{\rm gal}}(\ellB) = \hat\kappa^*_{\rm cmb}
(\ellB)\hat\kappa_{\rm gal}(\ellB) \,,
\end{eqnarray}
where the star denotes complex conjugation. Finally, we average over pixels in Fourier space with $|\ellB|\in (\ell-\Delta\ell/2, \,
\ell+\Delta\ell/2)$, for five linearly spaced bins between $40\leq\ell\leq2000$. 
We correct for the effect of the mask using an appropriate $f_{\rm sky}$ factor (including the
apodization), rather than computing and inverting the full mode-coupling matrix. Results from 
Ref.~\cite{Liu2014} and tests with simulations indicate that mask-induced effects only impact the power spectrum at $\ell > 7000$,
whereas we restrict our measurement to $\ell \leq 2000$ here (due to the band-limited Planck lensing maps).

To estimate the covariance matrix, we cross-correlate the CFHTLenS galaxy lensing maps with
100 simulated Planck CMB lensing maps (for both the 2013 and 2015 data, separately). We process these simulated maps through the same
pipeline as the actual Planck lensing maps. We then compute the covariance matrix from the
100 cross-power spectra. The diagonal components of the covariance matrix agree
to within 10\% with the theoretical variance
estimated using the auto-power spectra of the Planck and CFHTLenS maps (e.g., Eq.~30 of Ref.~\cite{Hill2014}). 
The off-diagonal components are relatively small, $\lesssim5\%$ of the diagonal terms.  We use the full
covariance matrices for all calculations in our results below.

\section{Results}\label{sec: results}

\subsection{Measurement}

\begin{figure}
\begin{center}
\includegraphics[scale=0.45]{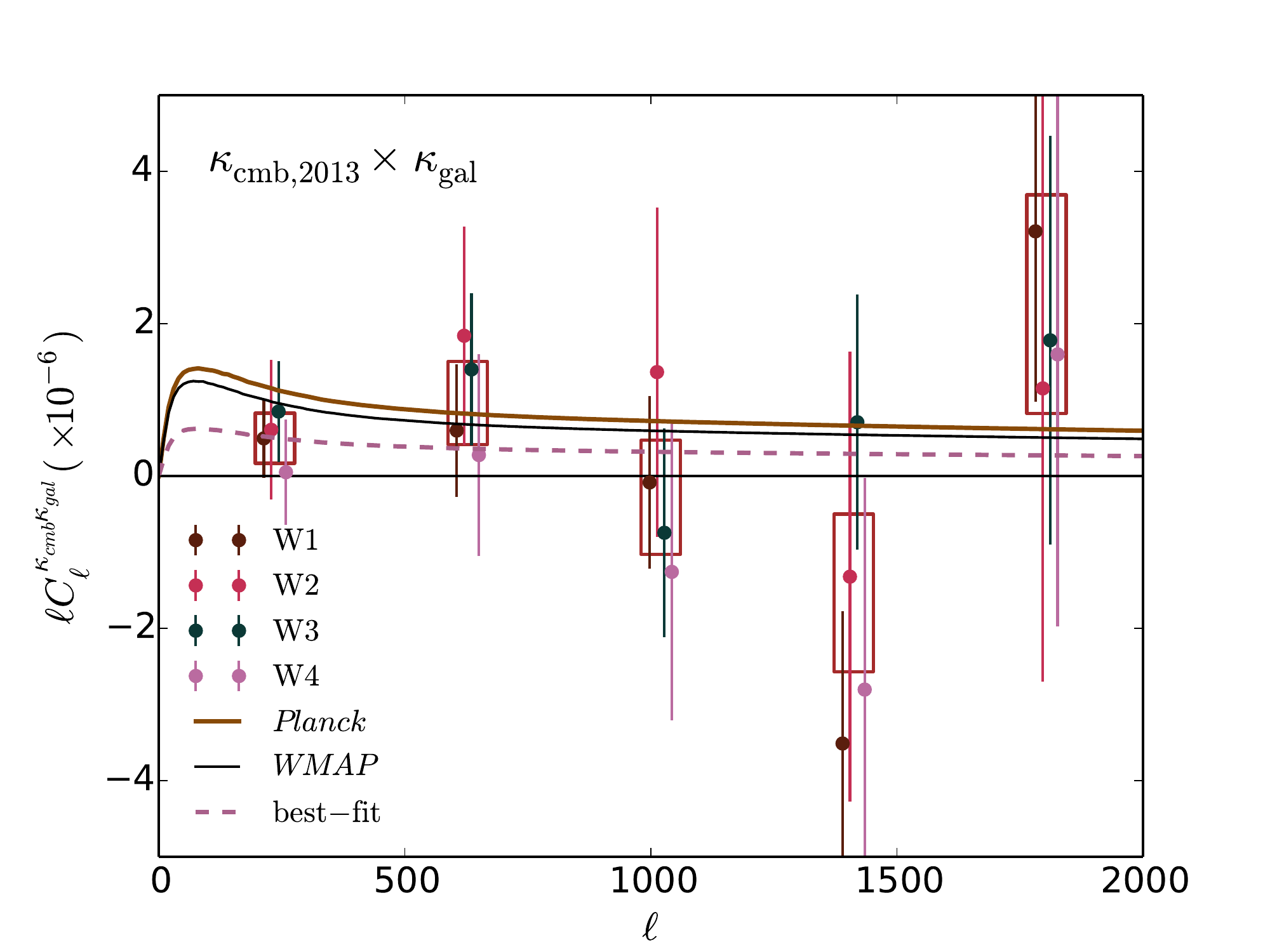}
\includegraphics[scale=0.45]{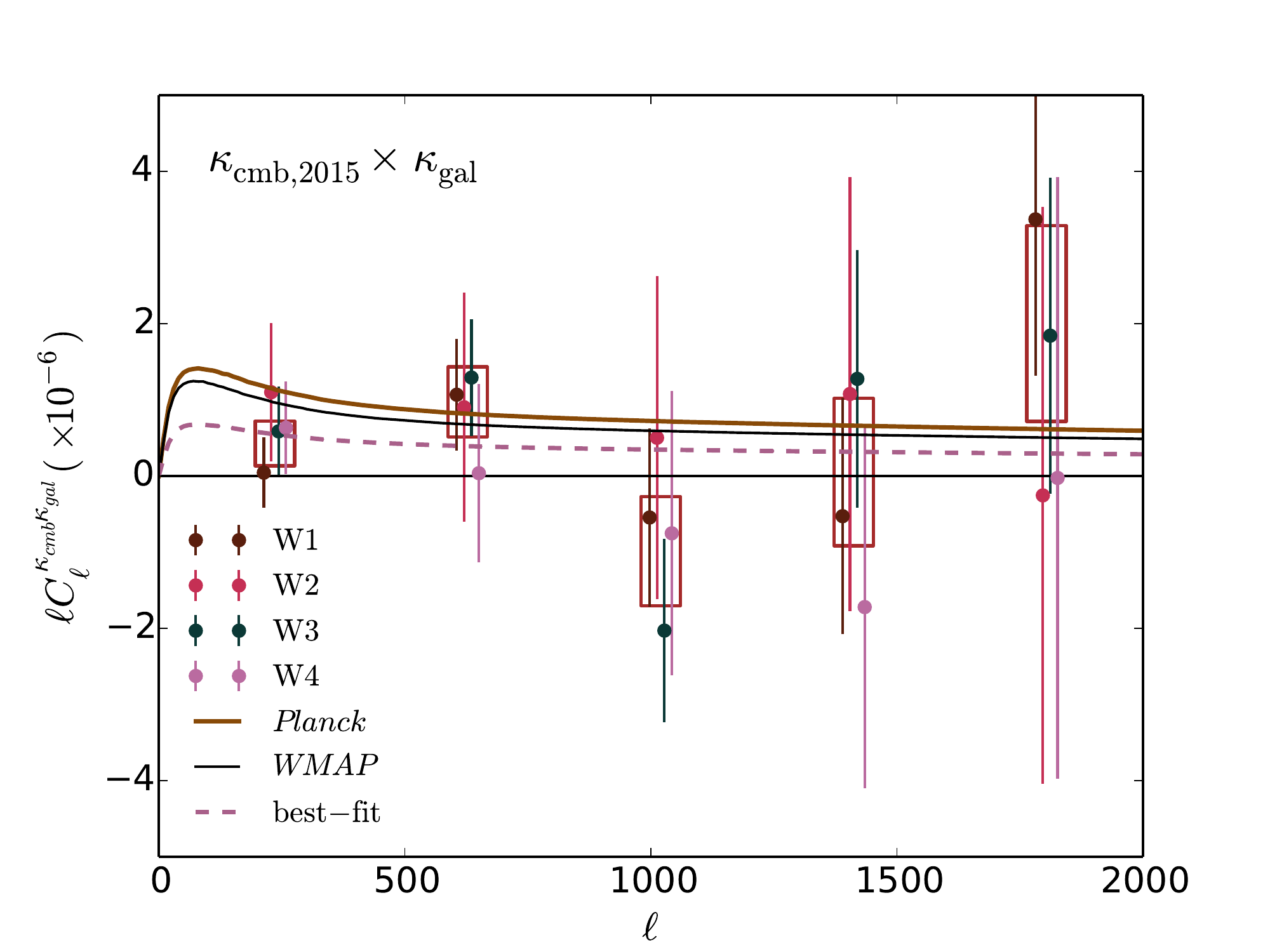}
\caption{\label{fig: CC} Cross-power spectra of Planck CMB lensing 
and CFHTLenS galaxy lensing maps. The top panel shows the result
for the 2013 Planck lensing map, while the bottom panel shows the 
2015 result. The solid curves are the (unscaled, i.e., $A=1$) 
theoretical prediction assuming Planck~\cite{Planck2015XIII} or WMAP
parameters~\cite{Hinshaw2013}.
The best-fit amplitudes with respect to the theory curves are $A_{2013}=0.48
\pm0.26$ and $A_{2015}=0.44\pm0.22$ using Planck 2015 parameters
(shown in dashed curves), 
and $A_{2013}=0.56 \pm0.30$ and $A_{2015}=0.52\pm0.26$ using WMAP9 parameters. 
Data points are for individual fields, and errors
are from the standard deviation of cross-power spectra between 100 simulated Planck
CMB lensing maps and CFHTLenS galaxy lensing maps. The boxes represent
the inverse-variance weighted sum of the four fields.}
\end{center}
\end{figure}

\begin{figure}
\begin{center}
\includegraphics[scale=0.45]{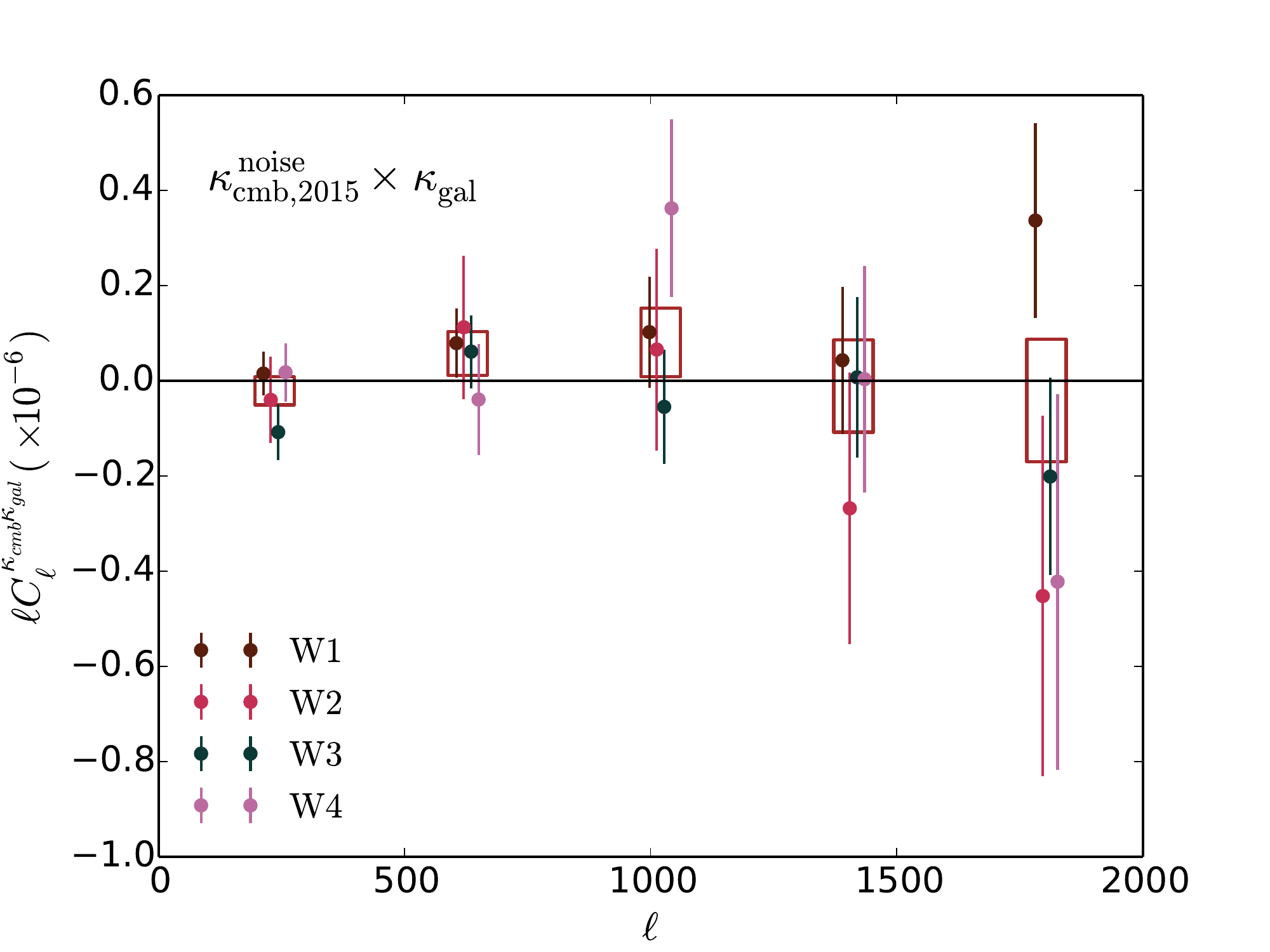}
\includegraphics[scale=0.45]{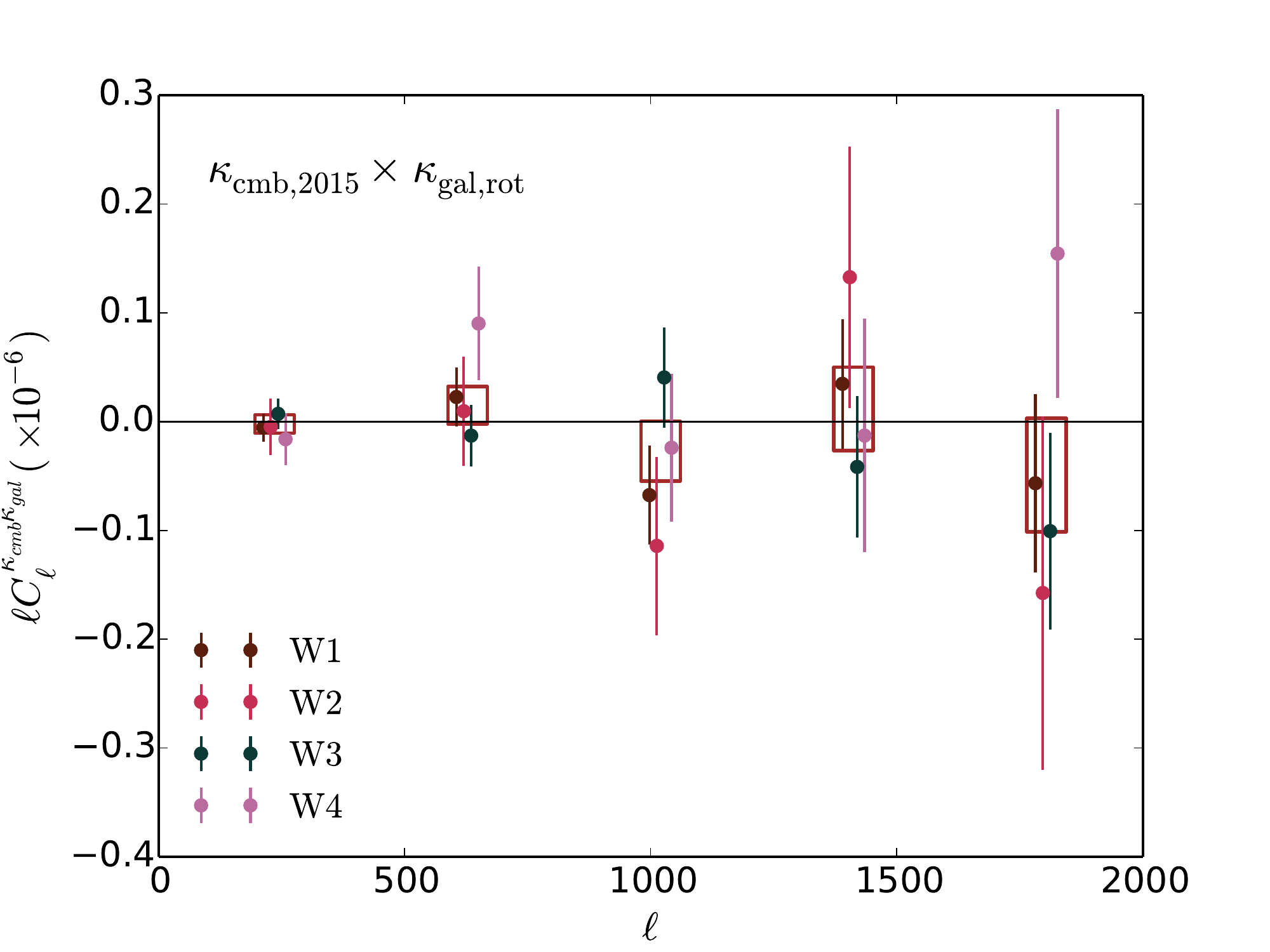}
\caption{\label{fig: CC_null} Null tests for the cross-power spectrum. In the upper panel,
we cross-correlate the CFHTLenS galaxy lensing maps and 100 simulated Planck
CMB lensing maps. In the lower panel, we cross-correlate the
Planck CMB lensing maps with 500 simulated galaxy lensing maps, obtained by randomly rotating
the CFHTLenS galaxies. Points are for individual fields, and errors
are the standard deviation of the simulated cross-power spectra divided by 
$\sqrt{N_{\rm sim}}$ with $N_{\rm sim}=100$ or 500 for the upper and lower panels, respectively. The boxes represent
the inverse-variance weighted sum of the four fields. We only show results using
the Planck 2015 CMB lensing maps. The results from the 2013 maps are also consistent with 
zero.}
\end{center}
\end{figure}

Fig.~\ref{fig: CC} shows the cross-power spectra of the Planck CMB lensing 
and CFHTLenS galaxy lensing maps. We use cosmological parameters from
either the Planck 2015~\cite{Planck2015XIII} or WMAP9 results~\cite{Hinshaw2013} to calculate
the theoretical prediction (shown as solid curves). We find the best-fit amplitude $A$ with respect
to the theoretical prediction by minimizing
\begin{eqnarray}
\label{eq: chisq}
\chi^2=\sum_{i,j}\left(C_{i}^{d}-AC_{i}^{m}\right) \CB_{ij}^{-1} \left(C_{j}^{d}-AC_{j}^{m}\right)
\end{eqnarray}
where $C^d$ is the cross-power spectrum calculated from data, $C^m$ is the
model calculated using Eq.~\ref{eqn:theory_eqn}, $i$ and $j$ denote the multipole bin (five bins for each CFHTLenS field), and
$\CB^{-1}$ is the inverse of the covariance matrix described above. The 
SNR is calculated as ${\rm SNR}=
\sqrt{\chi^2_{\rm null}-\chi^2_{\rm model}}$, where $\chi^2_{\rm null}=\chi^2(A=0)$ and $\chi^2_{\rm model}$
is the value for the best-fit amplitude $A$ (i.e., minimum $\chi^2$).

The best-fit amplitudes are shown in Table~\ref{tab: results}.
Using the 2013 Planck lensing map, we find $\chi^2_{\rm null, 2013}=19.1$
and $\chi^2_{\rm model, 2013}=15.6$ (for either Planck or WMAP parameters),
corresponding to SNR$=1.9$.  The probability-to-exceed (PTE) of the best-fit model is 
$0.68$. Using the 2015 Planck lensing map, we find $\chi^2_{\rm null, 2015}=17.2$,
and $\chi^2_{\rm model, 2015}=13.1$ (for either Planck or WMAP parameters),
corresponding to SNR$=2.0$.  The PTE of the best-fit model is 
$0.83$. In both cases, the model thus provides a good fit to the data. 

\begin{table}
\begin{tabular}{|c|c|c|}
\hline
	& $A$ (Planck parameters) 	&	$A$ (WMAP parameters)				\\
\hline															
2013 &	$0.48\pm0.26$	& $0.56\pm0.30$ \\
2015 &	$0.44\pm0.22$	& $0.52\pm0.26$\\
\hline
\end{tabular}
\caption[]{\label{tab: results}Best-fit amplitudes for the CMB lensing-galaxy 
lensing cross-power spectrum using Planck CMB lensing data (2013 and 2015 
releases, labeled by the rows) and the CFHTLenS galaxy lensing maps. The column
labels denote whether the amplitude $A$ is measured with respect to a theoretical model
computed with Planck 2015 cosmological parameters~\cite{Planck2015XIII} or WMAP9
parameters~\cite{Hinshaw2013}.}
\end{table}

To estimate constraints on cosmological parameters, we assume a power-law dependence
$C_{\ell}~\propto~(\sigma_8)^{x(\ell)} (\Omega_m)^{y(\ell)}$. Using the theoretical model discussed
in Sec.~\ref{sec: formalism}, we find that $x \approx 2$ in the 
linear regime ($\ell<$ few hundred) and $x \approx 3$ in the nonlinear regime ($\ell > 1000)$), with a gradual
transition in between. We find that $y\approx1.3$ for $\ell > 200$, and rapidly decreases to $y \approx -0.5$
at low-$\ell$. These power-law dependences are also apparent in Fig.~\ref{fig: model_varyingparams}.
We constrain a combination of parameters $\sigma_8(\Omega_m/0.27)^\alpha$,
which parametrizes the degeneracy between $\sigma_8$ and $\Omega_m$. For the cross-correlation
considered here, we find $\alpha=0.41$ for the best-constrained combination. Assuming a Gaussian likelihood, we obtain a best-fit
$\sigma_8(\Omega_m/0.27)^{0.41} = 0.63^{+0.14}_{-0.19}$. For reference, we also list
constraints from Planck primordial CMB measurements~\cite{Planck2013XVI,Planck2015XIII} and CFHTLenS cosmic shear
data~\cite{Heymans2013} in Table~\ref{tab: SIGMA8}. Our constraint
remains the same when using $\alpha=0.46$ for a direct comparison to Planck 2013 
and CFHTLenS. Our constraint is consistent with that from CFHTLenS, but
is in $\approx 2\sigma$ tension with Planck, as seen earlier in the best-fit amplitudes presented in Table~\ref{tab: results}.


We show two null test results in Fig.~\ref{fig: CC_null}, where we cross-correlate
(1) CFHTLenS galaxy lensing maps and 100 simulated Planck CMB lensing maps,
and (2) the CMB lensing maps and 500 simulated galaxy lensing noise maps, obtained by randomly rotating the CFHTLenS galaxies. 
The error bars are divided by $\sqrt {N_{\rm sim}}$ with the number of simulations
$N_{\rm sim}=100$ or 500 for the two cases, respectively. The results are consistent with zero, with PTE = 0.53 (2013 maps) and 0.11 (2015 maps) for test (1), and
PTE = 0.61 for test (2) for both Planck releases.

\subsection{Discussion}

The cross-correlation results present some puzzles. The SNR of the measurement ($\approx 2$) is substantially
below the predicted SNR $\approx 4.6$ computed in Sec.~\ref{sec: formalism}.  This result is entirely due to
the low amplitude of the measured signal with respect to the theoretical prediction.  The noise properties
are as expected --- the error bar on the measured amplitude for the Planck 2015 lensing -- CHFTLenS
cross-correlation agrees well with the forecast.  We find an error of $\sigma_A = 0.22$ (see Table~\ref{tab: results}), while
the prediction assuming $A=1$ is $\sigma_A = 0.217$.

Thus, the measured amplitude of the CMB lensing -- galaxy lensing cross-correlation
is in some tension with theoretical predictions using standard $\rm{\Lambda CDM}$.
The tension is most significant for the 2015 Planck lensing map, as seen in Table~\ref{tab: results}.
The measured amplitude ($A = 0.44 \pm 0.22$) in this case
is in tension with the prediction based on Planck 2015 cosmological parameters at the $2.5\sigma$
level. The tension is somewhat less significant for the 2013 lensing map ($2.0\sigma$), due to its
higher noise level and somewhat higher preferred amplitude ($A = 0.48 \pm 0.26$).  Note that the
decrease in amplitude from the 2013 to 2015 map ($\approx 10\%$) is responsible for the fact that the SNR hardly
improves when using the latter map, despite the lower noise (i.e., $\approx 15\%$ smaller error bar on $A$).
A similar amplitude shift from the 2013 to 2015 Planck lensing maps is reported in Ref.~\cite{Omori2015}, who use the same datasets
as in our analysis, but instead 
cross-correlate the Planck CMB lensing maps with the CFHTLenS galaxy number
density (rather than lensing convergence). They use the cross-correlation to infer the linear bias $b$
of the CFHTLenS galaxies, finding $b = 1.16^{+0.19}_{-0.18}$ for the 2013 release (for the $18.0 < i_{\rm AB} < 24.0$
CFHTLenS galaxy sample), but $b = 0.82^{+0.16}_{-0.14}$ for the 2015 release, a decrease of $\approx 30\%$. The shift
we observe is in the same direction as that in Ref.~\cite{Omori2015}, but at very low statistical significance. Note that
cross-correlation with CIB maps at 545 GHz in Ref.~\cite{Planck2015XV} does not show evidence of a significant shift in amplitude from
the 2013 Planck CMB lensing map to the 2015 map, which suggests that the small shift in our results is simply due to noise.

\begin{table}
\begin{tabular}{|c|c|c|c|c|c|}
\hline									
			&$\alpha$ &$\sigma_8(\Omega_m/0.27)^\alpha$&	Ref.	\\
\hline									
Planck 2013 CMB	&0.46	&$0.89^{+0.03}_{-0.03}$	&\cite{Planck2013XVI}	\\
Planck 2015 CMB	&0.50	&$0.90^{+0.02}_{-0.02}$	&\cite{Planck2015XIII}\\
CFHTLenS Cosmic Shear	&0.46	&$0.77^{+0.03}_{-0.04}$	&\cite{Heymans2013}	\\
This work		&0.41	&$0.63^{+0.14}_{-0.19}$	&--\\
\hline
\end{tabular}
\caption[]{\label{tab: SIGMA8} Cosmological parameter constraints.}
\end{table}

There are a number of potential reasons for the tension between our measured cross-correlation
amplitude and the $\Lambda$CDM prediction based on Planck 2015 cosmological parameters.
One possibility is that the true values of $\sigma_8$ and $\Omega_m$ are somewhat
lower than those found in the Planck CMB analysis. We note that the 
tension between our results and the theory is somewhat reduced when comparing to predictions based on WMAP9 cosmological
parameters (see the second column of Table~\ref{tab: results}). In this context, we refer the reader to the 
discussion in Ref.~\cite{Planck2015XIII} concerning
discrepancies between the Planck 2015 CMB-determined cosmological parameters and those determined
from CFHTLenS shear data (particularly $\sigma_8$ and $\Omega_m$). It is possible that modeling issues (e.g., the nonlinear power spectrum or $dn/dz$
uncertainties) affecting the weak lensing interpretation could be responsible, although
the lowest multipole bin in our measurement in Fig.~\ref{fig: CC} (where the theory is mostly in the linear regime) lies clearly below
the Planck 2015 theoretical prediction. Finally, we note that H15 also found a best-fit amplitude for the
ACT CMB lensing -- CS82 galaxy lensing cross-correlation that was slightly low compared to predictions based on
Planck cosmological parameters, though at smaller significance ($1.2\sigma$) than seen here. 

There are several systematics that could also be responsible for the
observed low amplitude. Photometric redshift uncertainties are an obvious suspect,
especially at high-$z$. We performed three tests in Sec.~\ref{sec: cfht} to assess the impact
of photo-$z$ uncertainties and found them
likely to be subdominant, but possibly on the order of $10$\%. With presently available data, 
we are unable to fully capture photo-$z$ systematics in galaxy spectral energy distribution modeling. 
To accurately quantify such uncertainties, one needs observations
extending further in the near infrared for high-redshift galaxies. Such a test is beyond the scope of this work.
Using an extreme test in which all $z>1.3$ galaxies in our data are moved to $z=1.3$, we find a rough upper
bound of $20$\% on photo-$z$ uncertainties in the cross-correlation amplitude. Given that the errors on the
high-redshift photo-$z$ are approximately symmetric~\cite{Ilbert2009}, this extreme test is likely an overestimate of the effect.
Further investigation in this area is clearly needed (as noted in H15).  We conclude that photo-$z$ errors alone are unlikely
to fully explain the observed low amplitude of the cross-correlation, but their effects are non-negligible ($\approx 10$\%).

A likely physical effect that contributes to the observed low amplitude is the intrinsic alignment~(IA)
of the foreground galaxy shape and the source shape distortion.
If a foreground galaxy is located between two overdense regions,
it can be tidally stretched in a direction perpendicular to the major axis of
the dark matter distribution. However, the shearing of source light will be 
aligned with the dark matter major axis, and hence the observed 
power spectrum amplitude will be reduced.
Refs.~\cite{Hall2014, Troxel2014} estimate the suppression due to this effect to be 
$\approx 15\%$ for CMB lensing -- galaxy lensing cross-correlations. To fully account for the $\approx 50\%$ difference 
seen in our results compared to standard $\Lambda{\rm CDM}$ solely with IAs, a very large IA amplitude would be necessary~($\approx3$ times
larger than the conservatively expected level). Thus, this effect alone is unlikely to fully explain the discrepancy, but could be non-negligible\footnote{Recently,
Ref.~\cite{Chisari2015} presented updated calculations of IA contamination for galaxy lensing -- CMB lensing cross-correlations, finding that
well-constrained low-redshift contributions were consistent with $10$--$20$\% contamination, but that unconstrained high-redshift
contributions could lead to an overall contamination as large as $60$\%.}.

Another possible contribution to the observed low amplitude relates to the mask used
in the construction of the Planck 2015 CMB lensing map. In the 2015 CMB lensing reconstruction, regions where 
tSZ clusters are located are masked prior to the reconstruction. The map thus contains no signal at these locations,
but because these clusters reside in overdense regions where lensing signals are expected, the mask
could affect our measurement. In contrast, the 2013 Planck CMB lensing reconstruction contains signal at the location
of tSZ clusters, because the analysis includes an independent reconstruction at 217 GHz, where the tSZ signal is null
and no cluster masking is required.  (The 2015 analysis is performed on a frequency-combined SMICA map, and thus
cluster masking is needed.) In our fiducial analysis above, we combined the 2013 and 2015 masks, and thus tSZ clusters
are masked. Since the 2015 lensing map simply does not include signal at the location of tSZ clusters, we cannot use it to
study the effect of this masking. However, the 2013 map does include such signal (because of the 217 GHz reconstruction),
and thus we can study the effect of the tSZ cluster mask by re-running our analysis on the 2013 map using only the 2013 lensing mask,
which does not cover tSZ clusters. 

Performing this analysis, we find $A=0.53 \pm 0.25$ compared to the model based on Planck 2015
parameters, and $A=0.62 \pm 0.30$ compared to the WMAP9 model. These amplitudes are $\approx 10$\% higher than those found 
using the 2015 mask which covers tSZ clusters ($0.48\pm0.26$ and $0.56\pm0.30$, respectively --- see Table~\ref{tab: results}).
The sky fraction in our analysis changes very little between the two masks: $f_{\rm sky,2013} / f_{\rm sky,2015} - 1 < 0.02$. Thus, the increased amplitude
is likely due to the inclusion of additional lensing signal at the location of the tSZ clusters. However, with only one
realization of the sky and a relatively noisy measurement, we cannot rule out the possibility that the increased amplitude
is simply a fluctuation due to including additional data. Testing this effect in a dedicated suite of simulations with correlated tSZ
and lensing signals is needed for a careful assessment. Also, note that because current CMB lensing auto-power spectrum
measurements are almost entirely in the linear regime, this effect is likely much smaller there than in the
cross-correlation with galaxy lensing studied here, which receives important nonlinear contributions over
most of the relevant multipole range (see Fig.~\ref{fig: model_haloterms}). However, this effect could be important for the galaxy number density -- CMB
lensing cross-correlation studied in Ref.~\cite{Omori2015}. It would not explain the difference
that they observe between the 2013 and 2015 Planck lensing maps, because they apply the 2015 (and 2013) Planck
lensing masks to both maps in their analysis. But this effect would bias their derived amplitudes low.
 We defer a careful assessment to future work, but the results above
suggest that this tSZ cluster mask systematic could explain part of the discrepancy of our measured amplitudes with respect to $\Lambda{\rm CDM}$ predictions.

It is also possible, though very unlikely, that the leakage of other secondary anisotropies into the CMB lensing map could play a role in our results, since these effects are correlated with the lensing field (e.g.,~\cite{Hill2014,Battaglia2014,Munshi2014,Planck2013XVIII}).  As noted in Sec.~\ref{sec: data}, biases in CMB lensing reconstruction due to tSZ or CIB leakage are small for an experiment with Planck's resolution and noise levels~\cite{vanEngelen2014b,Osborne2014}, even with no masking of clusters or CIB sources.  Since the 2015 Planck reconstruction uses the frequency-cleaned SMICA CMB map
and further masks the brightest tSZ clusters (as described above), such effects are additionally suppressed.  Moreover, most of the CIB emission comes from higher redshifts than those probed by the CFHTLenS lensing kernel, rendering it even less of a worry for our analysis.  CMB lensing maps can have residual kinematic SZ (kSZ) signals, as the kSZ effect has the same frequency dependence as the primordial CMB fluctuations.  However, this leakage vanishes to first order for the kSZ signal, since the line-of-sight velocity of the scattering electrons is equally likely to be positive or negative.  Thus, the lowest-order term that could affect our results is the kSZ${}^2$ -- weak lensing correlation, which is highly subdominant compared to the CMB lensing - weak lensing correlation that we measure (note that the kSZ signal alone is already a second-order effect).  The kSZ${}^2$ leakage into Planck CMB lensing maps was quantified in recent simulations performed in Ref.~\cite{Melin2015}, who found no evidence for an impact of the kSZ on cluster mass estimation using CMB lensing.  Overall, we find it highly unlikely that other secondary anisotropies have induced significant biases in our results.

Finally, it is possible that instrumental systematics could account for the low amplitude of our measured cross-correlation. Measurements of galaxy ellipticities are subject
to multiplicative and additive biases arising from the PSF and other effects that must be carefully calibrated. Painstaking analysis using the GREAT and SkyMaker
simulations is undertaken in Ref.~\cite{Miller2013} to perform this calibration for the CFHTLenS data.  While no significant
additive bias is found (and such a bias would be very unlikely to cross-correlate with the CMB lensing maps anyhow), a non-trivial
multiplicative bias on the measured ellipticities, $(1+m) \approx 0.9$--$0.95$, is measured using the simulations.  The
bias is larger for low-SNR galaxies (i.e., low $w_i$ in the notation of Sec.~\ref{sec: cfht}), which
are often high-$z$ galaxies (a fact of particular relevance for our study).  Moreover, the uncertainty on this multiplicative bias
correction is fairly large, with values $0.85 \lesssim (1+m) \lesssim 1.0$ consistent with the calibration over a wide range
of galaxy SNR and photometric redshift (see Fig.~12 in Ref.~\cite{Miller2013}).  The multiplicative bias propagates directly
to the shear and hence the convergence values, which scale as $1/(1+m)$. Thus, if the true value of $(1+m)$ is smaller than
found in Ref.~\cite{Miller2013}, the derived convergence values will increase, as will the amplitude of the convergence
auto-statistics and the CMB lensing cross-power spectrum studied here.  The authors of Ref.~\cite{Miller2013} note the possibility
that the galaxy models considered in their simulations might not be sufficient to capture
the true complexity of actual galaxies, which could lead to a systematic error in the calibration of $(1+m)$.
It is unlikely to be large enough to fully reconcile the discrepancy seen in our results with respect to the predictions,
but changes in the derived $(1+m$) values within the allowed range in Ref.~\cite{Miller2013} could produce $\approx 5$--$10$\%
changes in the measured cross-correlation amplitude.
Clearly, this effect also has important implications for the previously-discussed tension between the Planck 2015 CMB-determined cosmological parameters and those determined
from CFHTLenS shear data~\cite{Planck2015XIII}, especially since $(1+m)$ enters quadratically in the shear two-point statistics.

Fortunately, this hypothesis can be tested using existing data, as noted in earlier analyses~\cite{Vallinotto2012,Das2013}.
One can directly measure $(1+m)$ by computing the cross-correlation of 
(1) galaxy lensing maps with maps of galaxy number density (preferably using a spectroscopic sample), and (2) CMB lensing maps with the same 
galaxy number density maps. By taking
ratios of the measured cross-power spectra, one is left with only a factor of $(1+m)$ and a geometric factor arising from the
lensing kernels~\cite{Das2013}.  There is also no cosmic variance if one uses the same galaxy sample in both measurements.
Note that if our measured CMB lensing -- galaxy lensing cross-correlation had higher SNR, we could attempt such a calibration
directly~\cite{Vallinotto2012} (one could also split the galaxy data by the shear weight factor and look for a spurious dependence), but given the low SNR, a joint approach with galaxy number density maps seems more feasible. 
We leave an assessment of this calibration for future work.  The primary underlying assumption is that the
CMB lensing measurements are themselves free of a multiplicative bias (i.e., that the quadratic estimators have been properly
normalized).  The use of these methods will substantially tighten the cosmological constraints from upcoming weak 
lensing surveys~\cite{Vallinotto2012,Das2013}.

\section{Summary and outlook}\label{sec: summary}

Weak lensing of the CMB and galaxies has recently emerged as a powerful tool to 
constrain cosmological parameters. 
In this work, we cross-correlate Planck 2013 and 2015 CMB lensing maps with CFHTLenS
galaxy lensing maps, and detect a $2\sigma$ signal, despite an expected significance of $4.6\sigma$. Our best-fit amplitudes with respect
to the theoretical predictions are in $\approx 2$--$2.5 \sigma$ tension with standard $\rm{\Lambda CDM}$,
with $A_{2013}=0.48\pm0.26$ and $A_{2015}=0.44\pm0.22$ using Planck 2015 parameters. 
The tension is reduced if we assume WMAP9 parameters, with $A_{2013}=0.56 \pm0.30$ and $A_{2015}=0.52\pm0.26$.
A similar discrepancy (but with smaller significance) is also found by H15, where ACT CMB lensing maps
and CS82 galaxy lensing maps are used. We discuss possible sources of such power suppression,
including intrinsic alignments~($\approx15\%$) and masking of tSZ clusters in the CMB lensing reconstruction~($\lesssim 10\%$).
In addition, photometric redshift uncertainties could affect the cross-correlation at the $\approx 10$\% level.
It is possible that other systematics not yet accounted for could also play a role, such as the impact of nonlinear evolution or baryons on the
matter power spectrum, or an overall multiplicative bias in the CFHTLenS shear calibration. Taken together,
the combination of all of these systematic effects can perhaps explain the tension in our results with respect
to the Planck 2015 $\Lambda$CDM prediction. However, further detailed analysis is needed to understand
these effects at the required level of precision.

\begin{table}
\begin{tabular}{|c|c|c|c|c|c|}
\hline									
	&	$n_{\rm gal} \, [{\rm arcmin}^{-2}]$ 	&	$f_{\rm sky}$	&	Expected SNR	&	Ref.	\\
\hline									
DES	&	10	&	0.1	&	14.4	&	\cite{DES2005}	\\
HSC	&	20	&	0.048	&	15.5	&	\cite{HSC2010}	\\
Euclid	&	35	&	0.2	&	34.1	&	\cite{Euclid2009}	\\
LSST	&	40	&	0.25	&	39.5	&	\cite{LSSTSciBook2009}	\\
\hline
\end{tabular}
\caption[]{\label{tab: forecast} Weak lensing survey specifications and SNR forecasts.}
\end{table}

Due to the limiting size of CFHTLenS survey, less than $1$\% of the available 
Planck CMB lensing data are used in this work. Therefore, future improvement of the cross-correlation 
lies in larger galaxy weak lensing surveys, which fortunately are already ongoing.
For upcoming galaxy weak lensing surveys (the Dark Energy Survey~(DES),
Euclid\footnote{\url{http://sci.esa.int/euclid/}}, the Hyper Suprime-Cam Survey~(HSC), and
the Large Synoptic Survey Telescope\footnote{\url{http://www.lsst.org/}} (LSST)), we estimate the SNR of the cross-correlation
with the Planck 2015 CMB lensing data using the methodology discussed in Sec.~\ref{sec: formalism}. 
For the CMB lensing auto-power spectrum, we use the signal and noise power spectra provided in the Planck 2015
release. For the galaxy weak lensing auto-power spectra, we consider only shape noise in addition to the cosmological
signal. The weak lensing survey specifications and SNR forecasts are shown in Table~\ref{tab: forecast}.  The source
galaxy redshift distributions match those used in~\cite{Battaglia2014}, 
except for DES, for which we use the redshift distribution given in~\cite{Lahav2010}.
The predictions are quite promising, with $\approx 15\sigma$ detections expected for DES and HSC,
and $\approx 35$--$40\sigma$ detections expected for Euclid and LSST.
The predicted SNR for the upcoming surveys will be even higher if one considers the CMB lensing
maps from Advanced ACT~\cite{Calabrese2014} or other high-resolution, ground-based CMB experiments.
Thus, these ongoing and future surveys will precisely measure the CMB lensing -- galaxy lensing
cross-correlation.  In addition, the comparison of results from multiple independent surveys will allow multiplicative
shear systematics to be overcome.  These measurements will definitively determine whether the current tension in our results with respect to 
Planck 2015 $\Lambda{\rm CDM}$ parameters is significant or simply a statistical fluctuation.

\begin{acknowledgments}

We thank the Planck and CFHTLenS teams for devoting enormous efforts to produce the public data used in this work. We also thank Jim Bartlett, Zoltan Haiman, Nick Hand, Duncan Hanson, Alexie Leauthaud, Blake Sherwin, and David Spergel for useful discussions. JL is supported by National Science Foundation~(NSF) grant AST-1210877. JCH is supported by the Simons Foundation through the Simons Society of Fellows. Simulations for this work were performed at the NSF Extreme Science and Engineering Discovery Environment (XSEDE), supported by grant number ACI-1053575. Some of the results in this paper have been derived using the HEALPix package~\cite{Gorski2005}.

\end{acknowledgments}


\end{document}